\documentstyle[aps,epsf]{revtex}

\begin{document}
%\twocolumn[\hsize\textwidth\columnwidth\hsize\csname @twocolumnfalse\endcsname
\title{Low energy electronic states in spheroidal fullerenes}
\author{M. Pudlak$^a$, R. Pincak$^{a,b}$ and V.A. Osipov$^b$}
\address{
$^a$Institute of Experimental Physics, Slovak Academy of Sciences,
Watsonova 47,043 53 Kosice, Slovak Republic\\
$^b$Joint Institute for Nuclear Research, Bogoliubov Laboratory of
Theoretical Physics,
141980 Dubna, Moscow region, Russia\\
e-mail: pudlak@saske.sk, pincak@saske.sk, osipov@thsun1.jinr.ru }
\address{\em (\today)}
\preprint \draft
\maketitle
\begin{abstract}
The field-theory model is proposed to study the electronic states
near the Fermi energy in spheroidal fullerenes. The low energy
electronic wavefunctions obey a two-dimensional Dirac equation on
a spheroid with two kinds of gauge fluxes taken into account. The
first one is so-called K spin flux which describes the exchange of
two different Dirac spinors in the presence of a conical
singularity. The second flux (included in a form of the Dirac
monopole field) is a variant of the effective field approximation
for elastic flow due to twelve disclination defects through the
surface of a spheroid. We consider the case of a slightly
elliptically deformed sphere which allows us to apply the
perturbation scheme. It is shown exactly how a small deformation
of spherical fullerenes provokes an appearance of fine structure
in the electronic energy spectrum as compared to the spherical case.
In particular, two quasi-zero modes in addition to the true
zero mode are predicted to emerge in spheroidal fullerenes. An
additional 'hyperfine' splitting of the levels (except the
quasi-zero-mode states) is found.

\end{abstract}

\vskip 0.1cm \pacs{PACS numbers: 73.22.-f, 71.20.Tx, 73.21.-b}
\vskip 0.1cm

%The energy shift of the electronic spectra of fullerenes with
%elliptical geometries within a gauge field-theory were predicted.

\section{Introduction}

Geometry, topological defects and the peculiarity of graphene
lattice have a pronounced effect on the electronic structure of
fullerene molecules. The most extensively studied $C_{60}$ molecule
is an example of a spherical fullerene nicknamed a "soccer
ball"~\cite{Kroto}. The family of icosahedral spherical fullerenes is
described by the formula $20(n^2+nl+l^2)$ with integer $n$ and $l$.
Other fullerenes are either
slightly (as $C_{70}$) or remarkably deformed and their general
nickname is a "rugby ball". The electronic structure of $C_{70}$
cluster of $D_{5h}$ geometry has been studied in~\cite{Saito} by
using the local-density approximation in the density-functional
theory. It was clearly shown that the spheroidal geometry is of a
decisive importance for observed peculiarities in electronic states
of the $C_{70}$. What is important, the calculated energy levels
were found to be in good agreement with photoemission experiments.

Among different theoretical approaches to describe graphene
compositions such as nanotubes, cones, nanohorns and fullerenes
the continuum models play a special role. Indeed, the continuum
models allow us to describe some integral features of similar carbon
structures. For instance, in the case of the spherical fullerenes
the obtained results can be extended on the whole $60n^2$ family.
On the other hand, the continuum description gives an important information
about long-distance physics which is difficult to get within
other approaches like, e.g., widely-used tight-binding models. In
addition, the continuum models are useful for clarification of pure
topological effects such as an appearance of the Aharonov-Bohm
phase and anomalous Landau levels (see, e.g., Refs.~\cite{Crespi,Ando1,Furtado}).

It should be noted that the formulation of any continuum model
begins from the description of a graphite sheet (graphene)
(see, e.g.,~\cite{Wallace,Lomer,Weiss,Luttinger,DiVincenzo,Kane}).
The models which involve the influence of both geometry and topological
defects on the electronic structure by boundary conditions were
developed in Refs.~\cite{Ando,Dresselhaus}. A different variant of the
continuum model within the effective-mass description for fullerene
near one pentagonal defect was suggested in Refs.~\cite{Kochetov,Pudlak}.
There are attempts to describe electronic structure of the graphene
compositions by the Dirac-Weyl equations on the curved surfaces where
the structure of the graphite with pentagonal rings taken into account is imposed
with the help of the so-called K spin contribution~\cite{Guinea,Crespi}.
In the case of spherical geometry, this is realized by introducing an effective
field due to magnetic monopole placed at the center of a sphere~\cite{Guinea}.
In the present paper we use a similar approach.
The difference is that we consider also the elastic
contribution. Indeed, pentagonal rings being the disclination
defects are the sources of additional strains in the hexagonal
lattice. Moreover, the disclinations are topological defects and the
elastic flow due to a disclination is determined by the topological
Frank index. For this reason, this contribution exists even within
the so-called "inextensional" limit which is rather commonly used in
description of graphene compositions (see, e.g.,~\cite{Tersoff}).

Recently, in the framework of a continuum approach the exact
analytical solution for the low energy electronic states in
icosahedral spherical fullerenes has been found~\cite{Osipov}. The
case of elliptically deformed fullerenes was studied
in~\cite{Pincak} where some numerical results were presented. In
this paper, we suggest a similar to~\cite{Osipov} model with the
Dirac monopole instead of 't Hooft-Polyakov monopole for describing elastic
fields. We consider a slightly elliptically deformed sphere with the
eccentricity of the spheroid $e\ll 1$. In this case, by analogy
with~\cite{Clemenger} we use the spherical representation for the
eigenstates, with the slight asphericity considered as a
perturbation. This allows us to find explicitly the low-lying
electronic spectrum for spheroidal fullerenes.

\section{The model}

Our studies cover spherical molecules which are slightly
elliptically deformed. Let us start with introducing spheroidal
coordinates and writing down the Dirac operator for free massless
fermions on the Riemannian spheroid $S^{2}$. The Euler's theorem
for graphene requires the presence of twelve pentagons to get the
closed molecule. In a spirit of a continuum description we will
extend the Dirac operator by introducing the Dirac monopole field
inside the spheroid to simulate the elastic vortices due to twelve
pentagonal defects. The K spin flux which describes the exchange
of two different Dirac spinors in the presence of a conical
singularity will also be included in a form of t'Hooft-Polyakov
monopole.

To incorporate fermions on the curved background, we need a set of
orthonormal frames $\{e_{\alpha}\}$, which yield the same metric,
$g_{\mu\nu}$, related to each other by the local $SO(2)$ rotation,
$$e_{\alpha}\to e'_{\alpha}={\Lambda}_{\alpha}^{\beta}e_{\beta},\quad
{\Lambda}_{\alpha}^{\beta}\in SO(2).$$ It then follows that
$g_{\mu\nu} = e^{\alpha}_{\mu}e^{\beta}_{\nu} \delta_{\alpha
\beta}$ where $e_{\alpha}^{\mu}$ is the zweibein, with the
orthonormal frame indices being $\alpha,\beta=\{1,2\}$, and the
coordinate indices $\mu,\nu=\{1,2\}$. As usual, to ensure that
physical observables are independent of a particular choice of the
zweibein fields, a local $so(2)$ valued gauge field $\omega_{\mu}$
is to be introduced. The gauge field of the local Lorentz group is
known as a spin connection. The Dirac equation on a surface
$\Sigma$ in the presence of the abelian magnetic monopole field
$A_{\mu}$ is written as~\cite{Davies}
\begin{equation}
i\gamma^{\alpha}e_{\alpha}^{\ \mu}[\nabla_{\mu} - iA_{\mu}]\psi =
E\psi, \label{eq:1}
\end{equation}
where $\nabla_{\mu}=\partial_{\mu}+\Omega_{\mu}$ with
\begin{equation}
\Omega_{\mu}=\frac{1}{8}\omega^{\alpha\ \beta}_{\ \mu}
[\gamma_{\alpha},\gamma_{\beta}] \label{eq:2},
\end{equation}
being the spin connection term in the spinor representation.

The elliptically deformed sphere or a spheroid
\begin{equation}
\frac{x^{2}}{a^{2}}+\frac{y^{2}}{a^{2}} +\frac{z^{2}}{c^{2}}=1,
\label{eq:3}
\end{equation}
may be parameterized by two spherical angles $q^{1}=\phi$,
$q^{2}=\theta$ that are related to the Cartesian coordinates $x$,
$y$, $z$ as follows
\begin{equation}
x=a~\sin\theta \cos\phi; \quad y=a~\sin\theta \sin\phi; \quad
z=c~\cos\theta.
\label{eq:4}
\end{equation}
The nonzero components of the metric tensor for spheroid are
\begin{equation}
g_{\phi\phi}=a^{2}\sin^{2}\theta; \quad
g_{\theta\theta}=a^{2}\cos^{2}\theta+c^{2}\sin^{2}\theta, \quad
\label{eq:5}
\end{equation}
where~$a,c\geq0,~0\leq\theta\leq\pi,~0\leq\phi<2\pi$.
Accordingly, orthonormal frame on spheroid is
\begin{equation}
e{_{1}}=\frac{1}{a~\sin\theta}~\partial_{\phi};\quad
e{_{2}}=\frac{1}{\sqrt{a^{2}\cos^{2}\theta+c^{2}\sin^{2}\theta}}~\partial_{\theta}
\label{eq:6}~,
\end{equation}
and dual frame reads
\begin{equation}
e{^{1}}=a~\sin\theta~d\phi;\quad
e{^{2}}=\sqrt{a^{2}\cos^{2}\theta+c^{2}\sin^{2}\theta}~d{\theta}\label{eq:7}
~.
\end{equation}
A general representation for zweibeins is found
to be
\begin{equation}
e{^{1}}_{\phi}=a~\sin\theta;\  e^{1}_{\ \theta}=0;\ e^{2}_{\
\phi}=0;\
e{^{2}}_{\theta}=\sqrt{a^{2}\cos^{2}\theta+c^{2}\sin^{2}\theta}
\label{eq:8}
~.
\end{equation}
Notice that $e^{\mu}_{\ \alpha}$ is the inverse of
$e^{\alpha}_{\ \mu}$. The Riemannian connection with
respect to the orthonormal frame is written as~\cite{Nakahara,Gockeler}
\begin{equation}
de^{1}=-\omega^{1}_{\ 2}\wedge e^{2}=\frac{a
\cos\theta}{\sqrt{a^{2}\cos^{2}\theta+c^{2}\sin^{2}\theta}}\
d\phi\wedge e^{2},
\label{eq:9}
\end{equation}
\begin{equation}
de^{2}=-\omega^{2}_{\ 1}\wedge e^{1}=0.
\label{eq:10}
\end{equation}
Here $\wedge $ denotes the exterior product and $d$ is the
exterior derivative. From Eqs.(9) and (10) we get the Riemannian connection
in the form
\begin{equation}
\omega{_{\phi 2}^{1}}=-\omega{_{\phi 1}^{2}}=\frac{a
\cos\theta}{\sqrt{a^{2}\cos^{2}\theta+c^{2}\sin^{2}\theta}}\ ; \quad
\omega{_{\theta 2}^{1}}=\omega{_{\theta 1}^{2}}=0.
\label{eq:11}
\end{equation}
We assume that the eccentricity of the spheroid
$e=\sqrt{|1-(c/a)^{2}|}$ is small enough. In this case, one can
write down $c=a+\delta a$ where $\delta$ ($\mid\delta\mid\ll1$) is
a small dimensionless parameter characterizing the spheroidal
deformation from the sphere. So, we can follow the perturbation
scheme using $\delta$ as the perturbation parameter.

Within the perturbation scheme, the spin connection coefficients
are rewritten as
\begin{equation}
\omega{_{\phi 2}^{1}}=-\omega{_{\phi
1}^{2}}\approx\cos\theta(1-\delta\sin^{2}\theta).
\label{eq:12}
\end{equation}

The inclusion of the K-spin connection can be performed by
considering two Dirac spinors as the two components of an $SU(2)$
color doublet $\psi= (\psi_\uparrow \psi_\downarrow)^T$ (see
Ref.~\cite{Guinea} for details). In this case, the interaction with
the color magnetic fields in a form of nonabelian magnetic flux ('t
Hooft'-Polyakov monopole) is responsible for the exchange of the two Dirac
spinors which in turn corresponds to the interchange of Fermi
points. As was shown in Ref.~\cite{Gonzales}, the Dirac equation for
$\psi$ can be reduced to two decoupled equations for $\psi_\uparrow$
and $\psi_\downarrow$, which include an Abelian monopole of opposite
charge, $g=\pm 3/2$.

The elastic flow through a surface due to a disclination has a
vortex-like structure~\cite{Kochetov} and can be described by
Abelian gauge field $W_\mu$. Similarly to Ref.~\cite{Guinea}, we
replace the fields of twelve disclinations by the effective field
of the magnetic monopole of charge $G$ located at the center of
the spheroid.
%The vector potential of magnetic monopoles was
%chosen with nodal line runs from the origin along the line
%$\theta=\pi$.
The circulation of this field is determined by the
Frank index, which is the topological characteristic of a
disclination. Trying to avoid an extension of the group we
will consider the case of the Dirac monopole. The another
possibility is to introduce the non-Abelian 't Hooft-Polyakov
monopole (see Ref.~\cite{Osipov}). As is known, the vector
potential $W_\mu$ around the Dirac monopole have singularities. To
escape introducing singularities in the coordinate system let us
divide the spheroid (similarly as it was done for the sphere
in~\cite{Yang}) into two regions, $R_{N}$ and $R_{S}$, and define
a vector potential $A_{\mu}^{N}$ in $R_{N}$ and a vector potential
$A_{\mu}^{S}$ in $R_{S}$. Notice that $A_\mu$ includes both gauge
fields. One has

\begin{equation}
R_{N}:\ \ 0\leq\theta<\frac{\pi}{2}+\Delta;\ 0\leq\phi<2\pi,
\label{eq:13}\end{equation}
\begin{equation}
R_{S}:\ \ \frac{\pi}{2}-\Delta<\theta\leq\pi; \ 0\leq\phi<2\pi,
\label{eq:14}\end{equation}
\begin{equation}
R_{NS}:\ \ \frac{\pi}{2}-\Delta<\theta<\frac{\pi}{2}+\Delta;\
0\leq\phi<2\pi, \ (overlap) \label{eq:15}
\end{equation} where
$\Delta$ is chosen from the interval $0<\Delta\leq\pi/2$. In
spheroidal coordinates~(\ref{eq:5}) the only nonzero components
of $A_{\mu}$ are found to be
\begin{equation}
A^{N}_{\phi}\approx g\cos\theta(1+\delta\sin^{2}\theta) +
G(1-\cos\theta)-\delta G\sin^{2}\theta
\cos\theta,\label{eq:16}\end{equation}
\begin{equation}
 A^{S}_{\phi}\approx g\cos\theta(1+\delta\sin^{2}\theta)
 -G(1+\cos\theta)-\delta G\sin^{2}\theta \cos\theta,
\label{eq:17}
\end{equation}
where the terms of an order of $\delta^2$ and higher are dropped.
In the overlapping region the potentials $A_{\phi}^N$
and $A_{\phi}^S$ are connected by the gauge transformation
\begin{equation}
A_{\phi}^{N}=A_{\phi}^{S}+iS_{NS}\ \partial_{\phi}S^{-1}_{NS},
\label{eq:18}
\end{equation}
where
$$S_{NS}=e^{2iG\phi}$$
is the phase factor. The wavefunctions
in the overlap $R_{NS}$ are connected by
\begin{equation}
\psi_{N}=S_{NS}\psi_{S},
\label{eq:19}
\end{equation}
where $\psi_{N}$ and $\psi_{S}$ are the spinors in $R_{N}$ and $R_{S}$,
respectively. Since $S_{NS}$ must be single-valued~\cite{Wu},
$2G$ takes integer values. Notice that the total flux due to Dirac monopole
in (\ref{eq:16}) and (\ref{eq:17}) is equal to $4\pi G$.
What is important, there is no contribution from  the terms with $\delta$.
In our case, the total flux describes a sum of elastic
fluxes due to twelve disclinations, so that the total flux
(the modulus of the total Frank vector) is equal to $12\times \pi/3=4\pi$.
Therefore, one obtains $G=1$.

Let us consider the region $R_{N}$. The Dirac operator for spheroidal
fullerenes with monopole fields inside the spheroid takes the following form:
\begin{equation}
\hat{\cal{D}}=\hat{\cal{D}}_{0}+\delta\hat{\cal{D}}_{1},
\label{eq:21}\end{equation} where
$$\hat{\cal{D}}_{0}=i
\gamma_{2}\frac{1}{a}(\partial_{\theta}+\frac{\cot\theta}{2})+i\frac{\gamma_{1}}{a
\sin\theta}(\partial_{\phi}-iG(1-\cos\theta)-ig\cos\theta),
$$
is the Dirac operator for spherical fullerenes and
$$
\hat{\cal{D}}_{1}=
-i\gamma_{1}\frac{1}{a\sin\theta}\left [\frac{\cos\theta\sin^{2}\theta}{2}\gamma_{1}\gamma_{2}
+i \left(g-G\right)
\sin^{2}\theta\cos\theta\right ]-i\gamma_{2}\frac{\sin^{2}\theta}{a}\partial_{\theta},
$$
is the perturbation part of the spheroidal Dirac operator. This
operator is hermitian on the spheroid. Actually, we consider the
Dirac operator on the spheroid as a perturbation of the Dirac
operator on the sphere (cf.~\cite{Clemenger}). For this reason,
$\hat{\cal{D}}$ should be transformed into the operator which is
hermitian on the sphere. After straightforward calculations one
obtains that the form of the operator $\hat{\cal{D}}_{0}$ remains
the same while $\hat{\cal{D}}_{1}$ is written as
%$$\hat{\cal{D}}=i
%\gamma_{2}\frac{1}{a}\left(\partial_{\theta}+\frac{\cot\theta}{2}\right)+i\frac{\gamma_{1}}{a
%\sin\theta}\left(\partial_{\Phi}-i\frac{G}{2}(1-\cos\theta)-ig\cos\theta)\right)+
%$$
%\begin{equation}
%i\gamma_{1}\frac{\delta}{a}\sin\theta\left(\partial_{\Phi}-i\frac{G}{2}\right).
%\end{equation}
%We see that perturbation part of Dirac operator is transformed to the form
\begin{equation}
\hat{\cal{D}}_{1}=-\frac{\gamma_{1}}{a}\sin\theta\left(j-2m\cos\theta\right).
\end{equation}

Let us turn back to Eq.(\ref{eq:1}) and formulate the eigenvalue problem
for Dirac operator on the spheroid. Since a surface
is a two dimensional space, the Dirac matrices can be chosen to be the Pauli
matrices, $\gamma_1=-\sigma_2, \gamma_2=-\sigma_1$. We can restrict our
consideration to only one of spinors, say to $\psi_\uparrow$.
By using the substitution
\begin{equation}
\left(%
\begin{array}{c}
  \psi_{A} \\
  \psi_B \\
\end{array}%
\right) =\sum_j \frac{e^{i(j+G)\phi}}{\sqrt{2\pi}}\left(%
\begin{array}{c}
  u_j(\theta) \\
  v_j(\theta) \\
\end{array}%
\right) ,j=0,\pm 1,\pm 2,\ldots \end{equation}
we obtain the Dirac equations for the spinor functions $u_j$ and $v_j$ in the form
\begin{equation}
\left(-i\sigma_{1}\frac{1}{a}(\partial_{\theta}+\frac{\cot\theta}{2})+\frac{\sigma_{2}}{a
\sin\theta}\left(j-m\cos\theta\right)+\delta
\hat{\cal{D}}_{1}\right) \left(\begin{array}{c}
  u_j(\theta) \\
  v_j(\theta) \\
\end{array}
\right)=E\left(\begin{array}{c}
  u_j(\theta) \\
  v_j(\theta) \\
\end{array}
\right), \label{eq:22}
\end{equation}
where $m=g-G$. To first order in $\delta$, the square of the spheroidal Dirac operator is written as
$
\hat{\cal{D}}^{2}=(\hat{\cal{D}}^{2}_{0}+\delta\hat\Gamma),
$
where
$
\hat\Gamma=(\hat{\cal{D}}_{0}\hat{\cal{D}}_{1}+\hat{\cal{D}}_{1}\hat{\cal{D}}_{0}).
$
In an explicit form
\begin{equation}
a^{2}\hat{\cal{D}}^2_{0}=-\frac{1}{\sin\theta}\partial_{\theta}\sin\theta\partial_{\theta}+\frac{1}{4}
+\frac{1}{4\sin^{2}\theta}+\sigma_{3}\frac{m-j\cos\theta}{\sin^{2}\theta}
+\frac{(j-m\cos\theta)^{2}}{\sin^{2}\theta},
\label{eq:23}\end{equation} and
\begin{eqnarray}
a^{2}\hat\Gamma=2j^{2}-jx\left(\sigma_{3}-6m\right)
-4m(m-\sigma_{3})x^{2}-2m\sigma_{3}. \label{eq:24}
\end{eqnarray}
Let us write the equation $\hat{\cal{D}}^2\psi=E^2\psi$ by using
the appropriate substitution $x=\cos\theta$ in Eqs.(\ref{eq:23}) and (\ref{eq:24}).
One obtains
\begin{eqnarray}
[\partial_x(1-x^2)\partial_x-\frac{(j-mx)^2-j\sigma_3
x+\frac{1}{4}+\sigma_3 m}{1-x^2}+\delta\hat\Gamma]\left(%
\begin{array}{c}
  u_j(x) \\
  v_j(x) \\
\end{array}%
\right)=-(\lambda^2-\frac{1}{4})\left(%
\begin{array}{c}
  u_j(x) \\
  v_j(x) \\
\end{array}%
\right), \label{25}
\end{eqnarray}
%where perturbation part has following form
%\begin{eqnarray}
%a^{2}\Gamma=-2j^{2}-jx\left(\sigma_{3}-2m\right).
%\label{eq:26}
%\end{eqnarray}
where $\lambda=aE$. When $\delta=0$, one has the
case of a sphere with magnetic monopole inside. In this case,
the exact solution is known (see, e.g.,~\cite{Osipov})
%\begin{eqnarray}
%\left[\partial_x(1-x^2)\partial_x-\frac{(j+mx)^2-j\sigma_3
%x+\frac{1}{4}-\sigma_3 m}{1-x^2}\right]\left(%
%\begin{array}{c}
%  u_{jn}^{0}(x) \\
%  v_{jn}^{0}(x) \\
%\end{array}%
%\right)=-(\lambda^2_{0n}-\frac{1}{4})\left(%
%\begin{array}{c}
%  u_{jn}^{0}(x) \\
%  v_{jn}^{0}(x) \\
%\end{array}%
%\right), \label{eq:27}\end{eqnarray}

$$
\left(%
\begin{array}{c}
  u_{jn}^{0} \\
  v_{jn}^{0} \\
\end{array}%
\right)
=\left(%
\begin{array}{c}
  C_{u}(1-x)^{\alpha}(1+x)^{\beta}P_{n}^{2\alpha,2\beta} (x) \\
  C_{v}(1-x)^{\mu}(1+x)^{\nu}P_{n}^{2\mu,2\nu} (x) \\
\end{array}%
\right),
$$
with the energy spectrum
\begin{equation}
(\lambda_{jn}^{0})^2=(n+|j|+1/2)^2-m^2. \label{eq:28}
\end{equation}
Here
\begin{eqnarray}
\alpha=\frac{1}{2}\left|j-m-\frac{1}{2}\right|,\beta=\frac{1}{2}\left|j+m+\frac{1}{2}\right|,\nonumber\\
\mu=\frac{1}{2}\left|j-m+\frac{1}{2}\right|,\nu=\frac{1}{2}\left|j+m-\frac{1}{2}\right|,
\label{eq:29}
\end{eqnarray}
$P_{n}^{2\alpha,2\beta} (x)$ and $P_{n}^{2\mu,2\nu} (x)$ are Jacobi
polynomials of $n$-th order, and $C_{u}$ and $C_{v}$ are the
normalization factors. These states are degenerate. For example, the
degeneracy of the zero mode is equal to six. Therefore, we have to
use the perturbation scheme for the degenerate energy
levels~\cite{Landau}. As a result, the energy spectrum for
spheroidal fullerenes is found to be
\begin{equation}
(\lambda_{jn}^{\delta})^2=(n+|j|+1/2)^2-m^2+\delta\big(2j^{2}+j
L^{1}+jL^{2}+L^{3}-2mD_{jn}\big). \label{eq:29a}
\end{equation}
Similarly to~\cite{Osipov}, the possible values of $j$ obey the
condition $|j|\geq||m|+1/2|$ for non-zero modes. Direct calculations
show that
\begin{equation}
L^{1}=-\frac{j(6m^{2}+1/2)}{p(p+1)}+\frac{jm^{2}(4m^{2}+5)}{p^{2}(p+1)^{2}}
,\label{eq:29a}
\end{equation}
\begin{equation}
L^{2}=-\frac{4jm
D_{jn}}{p(p+1)}+\frac{jm(8m^{2}+1)D_{jn}}{p^2(p+1)^{2}},
\end{equation}
\begin{eqnarray}
L^{3}=4m(m-1)|C_{u}|^{2}F_{n}(2\alpha,2\beta)I_{n}(2\alpha,2\beta)\\
\nonumber +4m(m+1)|C_{v}|^{2}F_{n}(2\mu,2\nu)I_{n}(2\mu,2\nu)
\end{eqnarray}
with $p=n+\beta+\alpha$,
\begin{eqnarray}
F_{n}(2\alpha,2\beta)=\frac{(n+1)(n+2\alpha+1)(n+2\beta+1)(n+2\alpha+2\beta+1)}
{(2p+1)(p+1)^{2}(2p-n+3)}\nonumber\\
+\frac{n(n+2\alpha)(n+2\beta)(n+2\alpha+2\beta)}{(2p-1)p^{2}(2p+1)}\nonumber,
\label{eq:30}\end{eqnarray}
\begin{eqnarray}
I_{n}(2\alpha,2\beta)=\frac{2^{2\alpha+2\beta+1}\Gamma(n+2\alpha+1)\Gamma(n+2\beta+1)}{n!(2p+1)\Gamma(n+2\alpha+2\beta+1)},
 \label{eq:30}\end{eqnarray}
\begin{equation}
D_{jn}=\frac{\Gamma(n+2\mu+1)\Gamma(n+2\nu+1)-\left(\frac{j+1/2}{j-m+1/2}\right)^{2}\frac{n+j-m+1/2}{n+j+m+1/2}
\
\Gamma(n+2\alpha+1)\Gamma(n+2\beta+1)}{\Gamma(n+2\mu+1)\Gamma(n+2\nu+1)+\left(\frac{j+1/2}{j-m+1/2}\right)^{2}\frac{n+j-m+1/2}{n+j+m+1/2}
\ \Gamma(n+2\alpha+1)\Gamma(n+2\beta+1)} \label{eq:32}\end{equation}
for $j>0$ and
\begin{equation}
D_{jn}=\frac{\Gamma(n+2\mu+1)\Gamma(n+2\nu+1)-\left(\frac{|j|+1/2+m}{|j|+1/2}\right)^{2}\frac{n+|j|-m+1/2}{n+|j|+m+1/2}
\
\Gamma(n+2\alpha+1)\Gamma(n+2\beta+1)}{\Gamma(n+2\mu+1)\Gamma(n+2\nu+1)+\left(\frac{|j|+1/2+m}{|j|+1/2}\right)^{2}\frac{n+|j|-m+1/2}{n+|j|+m+1/2}
\ \Gamma(n+2\alpha+1)\Gamma(n+2\beta+1)} \label{eq:30b}
\label{eq:33}\end{equation} for $j<0$. Finally, in the linear in
$\delta$ approximation, the low energy electronic spectrum of
spheroidal fullerenes takes the form
\begin{equation}
E_{jn}^{\delta}=E^0_{jn}+E^f_{jn}+E^h_{jn}
\label{eq:36}
\end{equation}
with
\begin{equation}
E^0_{jn}=\pm\sqrt{(2\xi+n)(2\eta+n)}, \quad E^f_{jn}=\frac{\delta
(2j^{2}+jL^{1})}{2E^0_{jn}}, \quad E^h_{jn}=\frac{\delta (j
L^{2}+L^{3}-2mD_{jn})}{2E^0_{jn}}, \label{eq:37}
\end{equation}
where $\xi=\mu\ (\nu)$ and $\eta=\beta\ (\alpha)$ for $j>0\  (j<0)$, respectively.
Here we came back to the energy variable $E=\lambda/a$ (in units of $\hbar V_F/a$
where $V_F$ is the Fermi velocity).

Notice that the non-diagonal matrix element of perturbation
$\langle e^{j\phi}|\langle\psi_{jn}|\Gamma|\psi_{-jn}\rangle|e^{-j\phi}\rangle$
turns out to be zero. Therefore, the states with opposite $j$ do
not mix and $j$ remains a good quantum number as would be
expected. As is seen from Eq. (\ref{eq:29a}), for $\delta=0$
both eigenstates and eingenvalues are the same as for
a sphere (cf. Ref.~\cite{Osipov}).
At the same time, Eqs. (\ref{eq:29a})--
(\ref{eq:30b}) show that the spheroidal deformation gives rise to
an appearance of fine structure in the energy spectrum.
The difference in energy between sublevels is found
to be linear in $\delta$ which resembles the Zeeman effect where
the splitting energy is linear in magnetic field.
In addition we found a further splitting of the states with opposite $j$.
The splitting is weakly pronounced ('hyperfine' splitting) and entirely
dictated by the topological defects. In other words, this splitting
is pure topological in its origin. The magnitude of hyperfine splitting
is determined by the topological "charge" $m$. In particular,
for $m=0$ (no defects) $E^h_{jn}=0$ and no splitting occurs.

Table 1 shows all three contributions (in compliance with Eq. (\ref{eq:36}))
to the first energy level for various fullerenes with different morphologies.
As is clearly seen, the first (double degenerate) level becomes shifted
due to spheroidal deformation. Schematically, the structure of the first energy
level is shown in Fig.1. The more delicate is the structure of the second level
which is presented in Table 2. In this case, the initial (for $\delta=0$)
degeneracy of $E^0_{jn}$ is equal to six. The spheroidal deformation provokes
an appearance of three shifted double degenerate levels (fine structure) which, in turn,
are splitted due to the presence of topological defects (see Fig.2).
Notice that the magnitude of both shifts and the splitting energy
increases with $j$ and $m$.
It should be stressed that the experimental observation of such
splitting would indicate the presence of topological
defects while a measured magnitude of the splitting gives
the value of the monopole charge. Notice that in our model we have two
possible charges: $m=g-G=\pm 3/2 -1$. For comparison, the model
suggested in~\cite{Gonzales,Guinea} predicts $m=\pm 3/2$. Thus,
the splitting energy could clarify the proper monopole charge due
to topological defects. Notice that a similar small splittings of the
spheroidal subshells due to $D_{5h}$ symmetry was discovered in
Ref.~\cite{Saito}.

Let us discuss briefly the zero-mode state. As stated above, for a
spherical fullerene there is a zero-mode state with a sixfold
degeneracy. For spheroidal fullerene, the zero-mode state also
exists, however, its degeneracy is twofold. In addition, there
appear two slightly shifted "quasi-zero" modes for $j=\pm1,\pm2$,
each of them is also twofold degenerate (there is no hyperfine
splitting for quasi-zero-mode states). Indeed, our study shows
that in this case $D_{jn}$ is equal to one, so that $A_{jn}$ in
(\ref{eq:29a}) becomes odd function of $j$. It should be stressed
that the similar conclusions follow from our analysis based on
spheroidal harmonics suggested in Ref.~\cite{Fackerell}. Thus, we
confirm the finding in~\cite{Saito} that there are only up to
twofold degenerate states in spheroidal fullerene. Generally, the
spheroidal deformation gives rise to a reduction in degeneracy of
energy levels in comparison with the spherical case.

%\newpage

%Finally, following perturbation theory one has
%\begin{equation}
%\lambda_{jn}^2=\left(\lambda_{0jn}+\delta\lambda_{1jn}\right)^{2}\approx
%\lambda_{0jn}^2+2\lambda_{0jn}\lambda_{1jn}\delta,
%\end{equation}
%where $\lambda_{1jn}$ is the energy shift by perturbation. From
%equation (32) we get
%\begin{equation}
%\lambda_{1jn}=\frac{2 j^{2}+jA_{jn}}{2 \sqrt{(n+|j|+1/2)^2-m^2}}.
%\end{equation}

\section{Conclusion}

We have considered the electronic states of spheroidal fullerenes
provided the spheroidal deformation from the sphere is small
enough. In this case, the spherical representation is used for
describing the eigenstates of the Dirac equation, with the slight
asphericity considered as a perturbation. The using of the
perturbation scheme allows us to find the exact analytical
solution of the problem. In particular, the energy spectrum of
spheroidal fullerenes is found to possess the fine structure in
comparison with the case of the spherical fullerenes. The energy
between sublevels is found to be linear in the small distortion
parameter $\delta$ and is positive/negative for prolate/oblate
spheroid, respectively. Another principal difference is the
obtained hyperfine splitting of the energy levels. We have shown
that this splitting is weakly pronounced and entirely dictated by
the topological defects, that is it has a topological origin. Our
finding confirms the results of~\cite{Saito} that there can exist
only up to twofold degenerate states in the $C_{70}$.

Notice that for spherical fullerenes ($\delta=0$) our results agree
with those found in~\cite{Osipov}. It is interesting that the predictions of
the continuum model for spherical fullerenes are in qualitative
agreement with tight-binding calculations~\cite{Manousakis,Tang,Perez,Lin}.
In particular, the energy gap between the highest-occupied and lowest-unoccupied
energy levels becomes more narrow as the size of fullerenes
becomes larger.
It is important to keep in mind, however, that the continuum model itself
is correct for the description of the low-lying electronic states.
In addition, the validity of the effective field approximation for
the description of big fullerenes is also not clear yet. Actually,
this approximation allows us to take into account the isotropic part
of long-range defect fields. For bigger fullerenes, one has to consider
the anisotropic part of the long-range fields, the influence of the short-range
fields due to single disclinations as well as the multiple-shell structure.
Therefore, we admit that the proper values of the energy levels can deviate
from our estimations.

Finally, the main results obtained for spheroidal fullerenes are the discovery of
(i) fine structure with a specific shift of the electronic levels upwards,
(ii) hyperfine splitting due to topological defects, and (iii)
three twofold degenerate modes near the Fermi level with one of them being the true
zero mode. In our opinion, these predictions are quite general for the fullerene
family and are of interest for experimental studies.

%\section{Appendix}
%Dirac operator on the spheroid can be expressed in the following
%form
%$$\hat{\cal{D}}=i
%\gamma_{2}\frac{1}{a}(\partial_{\theta}+\frac{\cot\theta}{2})+i\frac{\gamma_{1}}{a
%\sin\theta}(\partial_{\Phi}-i\frac{G}{2}(1-\cos\theta)-ig\cos\theta)
%$$
%\begin{equation}
%-\delta\left(
%i\gamma_{1}\frac{1}{a\sin\theta}[\frac{\cos\theta\sin^{2}\theta}{2}\gamma_{1}\gamma_{2}
%+i \left(g-\frac{G}{2}\right)
%\sin^{2}\theta\cos\theta]+i\gamma_{2}\frac{\sin^{2}\theta}{a}\partial_{\theta}\right),
%\end{equation}
%this operator is self-adjoint on the spheroid where
%\begin{equation}
%\sqrt{g}=a \sin\theta\left(1+\delta\sin^{2}\theta\right),
%\end{equation}
%here $g$ is the metric tensor determinant. We must transform above
%mentioned Dirac operator to the one self-adjoint on the sphere
%where
%\begin{equation}
%\sqrt{g}=a \sin\theta\ .\end{equation} After transformation we get
%$$\hat{\cal{D}}=i
%\gamma_{2}\frac{1}{a}\left(\partial_{\theta}+\frac{\cot\theta}{2}\right)+i\frac{\gamma_{1}}{a
%\sin\theta}\left(\partial_{\Phi}-i\frac{G}{2}(1-\cos\theta)-ig\cos\theta)\right)+
%$$
%\begin{equation}
%i\gamma_{1}\frac{\delta}{a}\sin\theta\left(\partial_{\Phi}-i\frac{G}{2}\right).
%\end{equation}

\vskip 0.2cm \vskip 0.2cm The work was supported in part by VEGA
grant 2/6193/26 of the Slovak Academy of Sciences, by the Science
and Technology Assistance Agency under contract No. APVT-51-027904
and by the Russian Foundation for Basic Research under Grant No.
05-02-17721.

\newpage
\begin{table}[htb]
\begin{center}
\begin{tabular}[textwidth]{l c c c c c c c c c}
\bfseries $n=0$,\quad $m=1/2$ & \bfseries
$\overline{b}({\textmd{\AA}})$& $\overline{R}({\textmd{\AA}})$&
$SD({\textmd{\AA}})$& $\delta$ & $j$ & \bfseries $
|E^{0}_{jn}| (eV)$
&\bfseries $|E^{f}_{jn}|(meV) $ & \bfseries $|E^{h}_{jn}|(meV)$\\
\hline \hline\bfseries C$_{180}$ &1.43
&6.129& 0.075&0.012&1&1.24&5&2.7\\
&&&&&-1&1.24&5&2.2\\
\hline \hline \bfseries S-C$_{240}$&1.43
&7.12& 0&0 &1&1.07&0&0\\
&&&&&-1&1.07&0&0\\
\bfseries YO-C$_{240}$&1.45
&7.03& 0.17&0.024 &1&1.094&9&4.6\\
&&&&&-1&1.094&9&3.8\\
\bfseries TI-C$_{240}$&1.46
&7.09& 0.18&0.025&1&1.092&10&5\\
&&&&&-1&1.092&10&4\\
\bfseries I-C$_{240}$&1.45
&7.27& 0.39&0.054&1&1.06&20&10\\
&&&&&-1&1.06&20&8\\
\hline \hline \bfseries S-C$_{540}$&1.41
&10.5& 0&0 &1&0.712&0&0\\
&&&&&-1&0.712&0&0\\
\bfseries I-C$_{540}$&1.43
&10.4& 0.52&0.05&1&0.729&13&6.4\\
&&&&&-1&0.729&13&5.2\\
\end{tabular}
\end{center}
\caption{{\footnotesize The structure of the first energy level
for various fullerenes with different morphologies given in Ref. [33]:
S -- spherical, I -- icosahedron faceted, TI -- truncated icosahedron;
except C$_{180}$ -- the structure given in Ref. [34] and
YO -- given in Ref. [35]. $\overline{b}$ is
average bond length, $\overline{R}\ (\overline{R}=a)$ is average radius, $SD$ is
standard deviation from a perfect sphere (see Refs. [33,34]), so that
$\delta=SD/\overline{R}$. The hopping integral is taken to be $t=2.5\ eV$ and
$V_F=3t\overline{b}/2\hbar$ .
}}
\label{tab}
\end{table}
\vskip 3cm

\begin{table}[htb]
\begin{center}
\begin{tabular}[textwidth]{l c c c c c c c c c}
\bfseries \bfseries YO-C$_{240}$& $\overline{b}={1.45\textmd{\AA}}$&
$\overline{R}={7.03\textmd{\AA}}$& $SD={0.17\textmd{\AA}}$&
$\delta=0.024$ & $j$ & \bfseries $ |E^{0}_{jn}| (eV)$
&\bfseries $|E^{f}_{jn}|(meV) $ & \bfseries $|E^{h}_{jn}|(meV)$\\
\hline\hline \bfseries $n=1$, $m=1/2$ &
&&&&1&1.89&6.5&2.5\\
&&&&&-1&1.89&6.5&2.3\\
\bfseries  $n=0$, $m=1/2$ &
&&&&2&1.89&26&2.2\\
&&&&&-2&1.89&26&1.8\\
\bfseries $n=0$, $m=-5/2$ &
&&&&3&1.89&4.6&13\\
&&&&&-3&1.89&4.6&18\\
\end{tabular}
\end{center}
\caption{{\footnotesize The structure of the second energy level
for YO-C$_{240}$.}}
\label{tab2}
\end{table}

\newpage
\begin{figure}[!ht]
\begin{center}
\epsfysize=4cm \epsffile{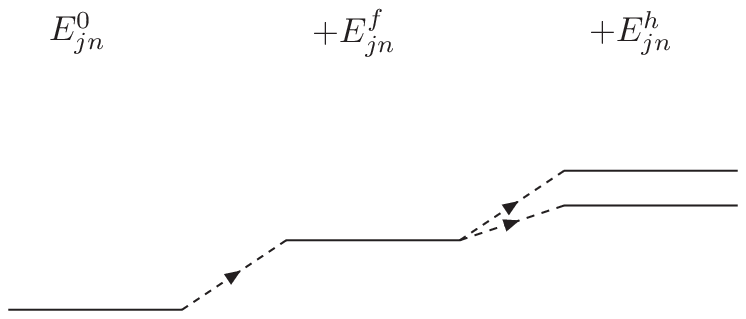}\epsfysize=4cm \caption[]{The
schematic picture of the first positive electronic level $E_{jn}^{\delta}$
for spheroidal fullerenes.}
%\label{DOS123}
\end{center}
\end{figure}
\vskip 3cm

\begin{figure}[!ht]
\begin{center}
\epsfysize=4cm \epsffile{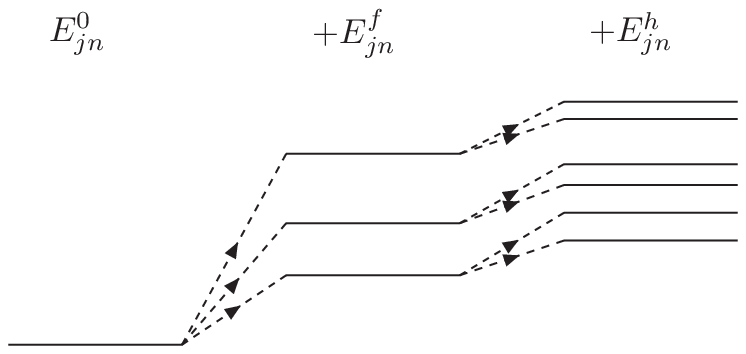}\epsfysize=4cm \caption[]{The
schematic picture of the second positive electronic level $E_{jn}^{\delta}$
for spheroidal fullerenes.}
%\label{DOS123}
\end{center}
\end{figure}

\end{document}